\documentclass[prb,aps,showpacs]{revtex4}
\usepackage{graphicx}
\usepackage{epstopdf}
\usepackage{amstext,amsmath,amsfonts,amsbsy,amssymb,bbm}
\usepackage{array,multirow}
\usepackage[pageshow]{supertabular}
\usepackage{color}
\newcommand {\be}[1]{\begin{eqnarray} \mbox{$\label{#1}$}  }
\newcommand{\ee}{\end{eqnarray}}

\newcommand{\pref}[1]{(\ref{#1})}

\newcommand{\nn}{\nonumber\\}
\newcommand{\noi}{\noindent}

\newcommand\hf{\frac 1 2 }
\newcommand\half{\frac 1 2 }

\newcommand{\dd}[2]{{d{#1}\over d{#2}}}
\newcommand{\ddtwo}[2]{{d^2{#1}\over d{#2}^2}}

\newcommand{\pdd}[2]{{\partial{#1}\over\partial{#2}}}

\newcommand{\pd}{\partial}

\newcommand{\ket}[1]{|#1\rangle}
\newcommand{\bra}[1]{\langle #1 |}

\newcommand{\gd}{ {\delta} }
\newcommand{\gD}{ {\Delta} }
\newcommand{\gk}{\kappa}
\newcommand{\gf}{\phi}
\newcommand{\gw}{ {\omega} }
\newcommand{\gq}{ {\theta} }
\newcommand{\gQ}{ {\Theta} }
\newcommand{\gy}{ {\psi} }
\newcommand{\gx}{ {\xi} }
\renewcommand{\ge}{ {\epsilon} }
\newcommand{\gl}{ {\lambda} }
\newcommand{\gr}{\rho}

\newcommand{\sgn}{{\rm sgn}}
\newcommand{\te}{{\tilde \ge}}
\newcommand{\tf}{{\tilde f}}
\newcommand{\brk}{{\bar k}}

\begin{document}

\title{Luttinger liquids, Fermi liquids and fractional statistics}
\author{Jon Magne Leinaas }
\affiliation{Department of Physics, University of Oslo, N-0316 Oslo, Norway}

\date{18 December 2016}
\begin{abstract}
We discuss how one-dimensional interacting fermion systems, which in the low energy approximation are described by Luttinger liquid theory, can be reformulated as systems of weakly interacting particles with fractional charge and statistics. Our approach is to use Landau's phenomenological approach to Fermi liquid theory, where the quasiparticles are interpreted as adiabatically dressed fermions. In an earlier publication the local charge carried by these excitations has been shown to be fractional. We focus here on the statistics of the quasiparticles and show that by a change of momentum variables the Landau parameters of the generalized Fermi fluid can be transformed to zero. This change in interaction is compensated by a change of the entropy function, which is consistent with the interpretation of the quasiparticles as satisfying generalized exclusion statistics.
\end{abstract}

\maketitle

\section{Introduction}
In one dimension the perturbative approach to Fermi liquid theory fails, in the sense that particle interactions generically give rise to intractable infrared divergences. However, Luttinger liquid theory, as shown by Haldane \cite{Haldane81}, can be viewed as a well functioning replacement. In this formulation bosonic operators, related to the particle densities, are the fundamental variables, and in the low-energy approximation, with linearized dispersion, the Hamiltonian takes a free-field form with the effects of interactions absorbed in velocity parameters. Non-linear corrections to the theory can in principle be included in the form of bosonic interaction terms.

Even if Luttinger liquid theory is well described in terms of bosonic variables, elementary charged excitations do exist in the one-dimensional Fermi system, and there has been in later years an interest in studying the properties of such excitations, both theoretically and experimentally \cite{FisherGlazman97, Pham00,Trauzettel04,LeHur08,Steinberg08,Berg09, LeinaasHorsdal09,Horsdal11,Inoue14}. The conclusion is that the excitations will generally carry a fraction of the fermionic charge. 
The fractionalization is linked to chiral separation of charges that are introduced in the system \cite{Safi95,Safi96}, so that fractions of a unit charge move to the right and the left, respectively. There has in particular been interest in the study of this effect for edge excitations in quantum Hall systems, where interactions between edge modes give rise to the charge fractionalization \cite{Berg09,LeinaasHorsdal09,Horsdal11,Inoue14}.

However, one should note the important difference between the charge fractionalization effect in the bulk of the quantum Hall system and at the edges. In the former case the excitations have well defined, topologically protected charge values, whereas in the latter case this is not the case. Instead the charges will there generally depend on the initial conditions which are responsible for the creation of the excitations. A particular way to introduce these excitations is to create them in the non-interacting system, and then adiabatically turn on the interaction between the edge modes.  This situation was examined in Refs.~\onlinecite{LeinaasHorsdal09} and \onlinecite{Horsdal11}, where it was shown how an integer charged edge excitation, in a quantum Hall bar with variable width, can adiabatically evolve into a fractionally charged excitation, with a compensating charge being smoothly absorbed into the background. 

In the present paper we focus on the {\em statistics} of the adiabatically dressed excitations. 
Our results are related to those in Refs.~\onlinecite{WuYu95, Isakov98,WuYu01}, where connections between generalized exclusion statistics and Luttinger liquids are discussed. However our approach, which is based on the use of Landau's phenomenological approach to Fermi liquid theory, is different. 
The starting point is the existence of an adiabatic mapping between the non-interacting and the interacting system of one-dimensional fermions, which makes the Fermi liquid theory valid. Since this mapping is known in the low energy approximation, the one-particle energy and the two-particle interactions can be derived from the energy of the system by functional differentiation with respect to the particle density. We further show, by use of a functional transformation, that the interaction terms  can be absorbed in such a way that the quantum statistics of the quasiparticles is effectively changed.

We will use in this paper the convention $\hbar=1$.

\section{The Luttinger liquid formulation}
We take as the starting point the following general expression for the Hamiltonian of a one-dimensional system of spinless fermions,
\be{ham1}
H=\sum_k\ge_0(k)c_k^\dag c_k+{1\over{4L}}\sum_{q,k_1,k_2}V(k_1-k_2,q)c_{k_1}^\dag c_{k_2}^\dag c_{k_2-q}c_{k_1+q}
\ee
$V$ is allowed to depend on the momentum variable $k_1-k_2$, in addition to $q$, in order to be able to include in this formulation the one-dimensional description of  quantum Hall systems with inter-edge interactions \cite{Horsdal07}.  Galilei invariance is not broken by the interaction, since it only depends on the relative variable $k_1-k_2$. 
We shall however assume that the dependence on this variable is week. This implies that we can disregard the effect of the $k$ dependence for low energy particles close to the same Fermi point, while for the interaction between particles at opposite Fermi points the effect will generally be significant.   For electrons in the lowest Landau level the condition of weak dependence means that the magnetic length is much smaller than the range of the interaction between the particles \cite{Horsdal07}.

The ground state of the Hamiltonian we assume to have the form of a filled Fermi sea, with well-defined Fermi points for negative and positive momenta, $k=\pm k_F$. In the low-energy approximation the excitations of the system are restricted to momenta near the two Fermi points, with the following low energy form of the Hamiltonian \cite{Haldane81},
\be{ham2}
H=  \bar v_F\sum_{\chi,k}(\chi k-k_F):c_{\chi,k}^\dag c_{\chi,k}: +{1\over{4L}}\sum_{\chi,q}[V_1(q)\gr_{\chi,q}\gr_{\chi,-q}+V_2(q)\gr_{\chi,q}\gr_{-\chi,-q}]
\ee
where $\chi=\pm 1$ is a chirality parameter, associated with the two Fermi points, and the operators being normal ordered with respect to the filled Fermi sea. The system is assumed to be confined to an interval of length $L$. With periodic boundary conditions for the fermion fields, the momentum then takes discrete values $k=2\pi n/L$, with $n$ as an integer. The interaction has been separated in two parts with $V_1(q)=V(0,q)$ and $V_2(q)=V(2k_F,q)$, and the density operators are normal ordered relative to the filled Fermi sea. The effective Fermi velocity $\bar v_F$ has the form
\be{Fermi}
\bar v_F=v_F-{1\over 4\pi }(V_1(0)-V_2(0))
\ee
with $v_F=\pdd{\ge_0}{k}(k_F)$ as the Fermi velocity of the non-interacting Fermi system, and the interaction dependent term is a correction, created by interactions between the low energy fermions and the Fermi sea \cite{Horsdal07}.

Although the $k$ quantum number is in the low energy approximation restricted to small deviations from $\pm k_F$, this restriction can be lifted, since the low energy sector of the theory is not affected by this extension. Without the restriction, the model \pref{ham2} describes in effect two types of fermions, characterized by different values of  $\chi$, both types with linear dispersion. 

The standard way to analyze the system described by the Hamiltonian \pref{ham2} is in terms of bosonization. We briefly summarize expressions to be used in the discussion to follow. 
The Fourier components of the charge density operators, for $q\neq 0$, are written as boson annihilation and creation operators as,
\be{aop}
a_q=\sqrt{{2\pi}\over |q|L}\sum_\chi \gq(\chi q)\,\rho_{\chi,\,q}\,,\quad 
a_q^\dag=\sqrt{{2\pi}\over |q|L}\sum_\chi \gq(\chi q)\,\rho_{\chi,\,-q}
\ee
with $\gq(q)$ as the Heaviside step function. The $q=0$ components of the charge densities define  the conserved fermion number and chiral (current) quantum number
\be{NJ}
N=\sum_{\chi}N_\chi=\sum_{k\, \chi}:c^\dag_{\chi,k} \,c_{\chi,k}:\,,\quad\quad J=\sum_{\chi}\chi N_\chi=\sum_{k\, \chi}\chi\, c^\dag_{\chi,k}\,c_{\chi,k}
\ee
with $N$ measuring the deviation of the particle number from its ground state value $N_0$. The bosonized form of the Hamiltonian is \cite{Haldane81}
\be{boseham}
H&=&{{\pi }\over{2L}}(v_N N^2+v_JJ^2)
\nn
&+&{ 1\over 2}\sum_{q\neq 0}|q|\left[
(\bar v_F+\frac{V_{1}(q)}{4\pi })
(a_q^{\dag}a_q+a_q a_{q}^{\dag})+\frac{V_{2}(q)}{4\pi }(a_q^{\dag}a_{-q}^{\dag}+a_q a_{-q})\right] 
\ee
which relative to the Hamiltonian \pref{ham2} has been modified by adding (non-relevant) terms that are constant or linear in $N$.  
The two velocity parameters $v_N$ and $v_J$ are 
\be{velpar}
v_N&=&\bar v_F+\frac{1}{4\pi }(V_1(0)+V_2(0))= v_F+\frac{1}{2\pi }V_2(0)\nn
v_J&=&\bar v_F+\frac{1}{4\pi }(V_1(0)-V_2(0))= v_F
\ee
We note in particular that $v_J$ is identical to the original Fermi velocity $v_F$ of the non-interacting fermions, rather than to the effective Fermi velocity $\bar v_F$ that appears in the Luttinger Hamiltonian \pref{ham2}. This can be viewed as a consequence of Galilei invariance of the original Hamiltonian \pref{ham1}.
The low energy sector, where \pref{boseham} is valid, corresponds to situations where $|q|$, as well as $N/L$ and $|J[/L$, are effectively restricted to values much smaller than $k_F$.

The bosonized Hamiltonian is diagonalized by a Bogoliubov transformation of the form
\be{bogo}
a_q=\cosh\xi_q\,b_q+\sinh\xi_q\,b^\dag_{-q}\nn
a^\dag_q=\cosh\xi_q\,b^\dag_q+\sinh\xi_q\,b_{-q}
\ee
where $\xi_q$ is fixed by the relation
\be{xi}
\tanh2\xi_q=-\frac{V_2(q)}{V_1(q)+4\pi  \,\bar v_F}
\ee
In terms of the new bosonic operators the Hamiltonian gets the diagonal form
\be{hamdiag}
H=\sum_{q\neq 0} \gw_q \,b_q^{\dag}b_q
+ {{\pi }\over{2L}}(v_N N^2+v_JJ^2)
\ee
with the frequency $\gw_q$ given by
\be{freq}
\gw_q=\sqrt{\left(\bar v_F+\frac{V_1(q)}{4\pi }\right)^2-\left(\frac{V_2(q)}{4\pi }\right)^2}\;|q|
\ee

The bosonized form of the low-energy Hamiltonian \pref{hamdiag} has for given $N$ and $J$ a free field form, which makes it straight forward to solve the many-particle problem and in particular to determine the relevant correlation functions \cite{Haldane81}.
However, for our purpose
it will be useful to reintroduce fermion variables in the expression for the Hamiltonian.

\section{Adiabatically dressed fermions}
The two sets of bosonic operators are unitarily equivalent,
\be{uni}
b_q=Ua_qU^\dag\,, b_q^\dag=Ua_q^\dag U^\dag
\ee
with the unitary transformation given by
\be{utrans}
U=\exp[-\sum_{q\neq 0}{\gx_q\over 2}(a_q^2-a_q^{\dag 2})]
\ee
The operator $U$ preserves the particle number of the two chiralities separately and it maps energy eigenstates of the linearized, free theory continuously into the eigenstates of the interacting theory, when the parameters $\gx_q$ are changed. The transformation can thus be interpreted as defining an adiabatic change of the energy eigenstates during a slow turning on of the interactions $V_1$ and $V_2$. 

For the fermion operators we introduce the corresponding transformation
\be{transfield}
\gf(x)=U\gy(x)U^\dag
\ee
and we refer to the states created by $\gf^\dag(x)$ as the dressed fermion states. In Ref.~\onlinecite{LeinaasHorsdal09} charged excitations created by the dressed fermion operator were studied, and it was shown that the integer fermion charge introduced by $\gf^\dag(x)$ was split in a non-integer, local contribution, equal to $\sqrt{g}=(v_J/v_N)^{1/4}$, and a compensating charge which was evenly distributed over the background.
 A further study of this effect in Ref.~\onlinecite{Horsdal11} confirmed this result numerically and showed how a slow change of the interaction would dynamically separate the total charge of a moving electron on the edge of the system into a sharply defined front pulse with charge $\sqrt{g}$  followed by a long, extended tail. 
 
 The demonstration of the charge fractionalization is in fact quite easy in the bosonic representation. To show this we focus on the Fourier components $\gr_q$ of the fermion density operator. For $q=0$ this is identical to the total fermion number, while the limit $q \to 0$ defines what is interpreted as the local part. The difference of these two is clearly seen in the case of a filled Fermi sea with $N$ particles, where $\gr_0= N$, while $\gr_q= 0$ for all $q\neq 0$. With excitations included, the charge component $\gr_q$ (for $q\neq 0$) has, in the low energy approximation, the following simple relation to the bosonic creation and annihilation operators
 \be{locch}
 \gr_q&=&\sqrt{\frac{L|q|}{2\pi}}(a_q+a_{-q}^\dag)\nn
 &=&\sqrt{\frac{L|q|}{2\pi}}(\cosh\xi_q+\sinh\xi_q)(b_q+b_{-q}^\dag)
 \ee
 which gives
 \be{qtran}
 U^\dag \gr_q U=(\cosh\xi_q+\sinh\xi_q) \gr_q
 \ee
 In the limit $q\to 0$ we have the following expressions for  $\cosh\xi_q$ and $\sinh\xi_q$,
 \be{cs}
\cosh\xi_0= \frac{g+1}{2\sqrt g}\,,\quad \sinh\xi_0=\frac{g-1}{2\sqrt g}
\ee
which gives
\be{lim}
\lim_{q\to 0}U^\dag \gr_q U=\sqrt g \lim_{q\to 0}\gr_q
\ee

Let us assume that $\Psi^\dag$ adds a local charge $Q=1$ to the Fermi sea of the noninteracting system, with the new (normalized) state written as
\be{crea}
\ket \psi =\Psi^\dag\ket F
\ee
$\ket F$ represents the filled Fermi sea, and the assumption about the charge is 
\be{ch}
\lim_{q\to 0} \bra F\Psi\gr_q\Psi^\dag\ket F=\bra F\Psi\gr_0\Psi^\dag\ket F=1
\ee
The dressed state is $U\Psi^\dag\ket F$, and the corresponding (local) dressed charge is then
\be{drech}
\lim_{q\to 0} \bra F\Psi U^\dag\gr_qU \Psi^\dag\ket F=\sqrt g \lim_{q\to 0} \bra F\Psi\gr_q\Psi^\dag\ket F=\sqrt g
\ee
which shows the result that, with repulsive interaction ($g<1$), only a part of the original fermion charge appears at the local charge of the dressed fermion, while the remaining part is distributed evenly as a part of the background.

To show the unconventional statistics of the dressed fermions is however not so straight forward.   The transformed field $\gf(x)$ clearly satisfies the same anticommutation relations as the electron field $\psi(x)$, and in this sense is a fermion field. However,  the statistics of the particles is not necessarily apparent in the commutation relations of the fields alone,  since the form of the Hamiltonian may reveal the  presence of  a "statistical interactions" between the particles. For this reason we will examine more closely the form of the Hamiltonian, when this is expressed in fermionic variables. 

To proceed we first make the low-energy approximation
\be{lowen}
V_1(q)\approx V_1(0)\,,\quad V_2(q)\approx V_2(0)
\ee
which for the boson frequency implies 
\be{lim2}
\gw_q\approx v_s|q|,\quad v_s=\sqrt{v_J v_N}
\ee
For the transformation $U$ this gives
\be{utrans2}
U\approx\exp[-\sum_{q\neq 0}{\gx_0\over 2}(a_q^2-a_q^{\dag 2})]\,, \quad \tanh\xi_0=\frac{g-1}{g+1}
\ee
which implies that  $U$, in this approximation, is uniquely determined by the interaction parameter $g=\sqrt{v_J/ v_N}$.

We now separate the Hamiltonian in two parts in the following way
\be{ham3}
H=U v_s(\sum_{q\neq 0} |q| a_q^\dag a_q+{{\pi }\over{2L}}(N^2+J^2))U^\dag
+v_s{{\pi }\over{2L}}(({1\over g}-1) N^2+(g-1) J^2)
\ee
where the first  term can be identified as a linearized free-field Hamiltonian, with $v_s$ as Fermi velocity, and with the field variables transformed by the operator $U$. This implies that the Hamiltonian can be expressed in terms of the dressed fermion field as
\be{ham4}
H=v_s 
\{
\int_0^L dx :\sum_\chi \phi_\chi^\dag(x)(-i\chi \pd_x - k_F)\phi_\chi(x):\nn
+{\pi\over{2L}}\sum_\chi[({1\over g}+g-2)N_\chi^2
+({1\over g}-g) N_\chi N_{-\chi}]\}
\ee
where the chiral fields are defined by
\be{chifield}
\psi_\chi(x)={1\over \sqrt{L}}\sum_k c_{\chi,k} e^{ikx}\,,\quad \phi_\chi(x)=U\psi_\chi(x)U^\dag
\ee

\vskip3mm
\noi
\section{Fermi liquid description}
In the low energy approximation we have an explicit expression \pref{utrans} for the operator $U$ that adiabatically changes the energy eigenstates of the non-interacting system into those of the interacting one. When corrections to this approximation are included, we may still assume the interacting system to be adiabatically connected to the non-interacting one, although there will be corrections to the form of the adiabatic mapping. 
This assumption of adiabatic connection between the non-interacting and interaction system forms the basis for Landau's  Fermi liquid theory, which therefore is applicable in the present case, beyond the low energy approximation \pref{ham4}.

The total energy is then given as a functional of the distribution of occupation numbers $n(k)$, associated with the non-interacting theory,
\be{efunk}
E = E[n(k)] 
\ee
and the quasiparticle energies and interactions can be defined in terms of functional derivatives to first and second order in the particle density \cite{Landau56},
\be{enpart}
\gd E=\sum_k \ge(k) \gd n(k)+\half\sum_{k k'} f(k,k')\gd n(k)\gd n(k')
\ee
The quasiparticles introduced by Landau in this way we will here identify with the dressed fermions previously discussed.
For variations about the filled Fermi sea, the expressions for energy and interactions will be referred to as  $\ge_0(k)$ and $f_0(k,k')$. 

In the low energy approximation the energy and interaction terms can be extracted from the Hamiltonian \pref{ham4}. With $\chi$ related to $k$ by $\chi=\sgn \,k$, the corresponding expression for the variation of the energy is
\be{varen}
\gd E=\sum_k v_s(|k|-k_F)\gd n(k)+v_s{\pi\over L}\sum_{k,k'}(\gl_1\gq(kk')+\gl_2\gq(-kk'))\gd n(k)\gd n(k')
\ee
with $\gq(k)$ as the Heaviside step function, and with $\gl_1$ and $\gl_2$ defined by
\be{lambda}
\gl_1=\half ({1\over g}+g-2)\,,\quad \gl_2=\half ({1\over g}-g)
\ee
From this follows that the single particle energy and the interaction terms are
\be{ef}
\ge_0(k)&=&v_s(|k|-k_F)\nn
f_0(k,k')&=&v_s{{2\pi}\over L}(\gl_1\gQ(kk')+\gl_2\gQ(-kk'))
\ee

In the case where corrections to the low energy Hamiltonian \pref{ham4} are included, we interpret the above expression for the interaction to be valid at the Fermi points, written as
\be{int}
f_0(k_F,k_F)&=&f_0(-k_F,-k_F)=v_s{{2\pi}\over L}\gl_1\nn
f_0(k_F,-k_F)&=&f_0(-k_F,k_F)=v_s{{2\pi}\over L}\gl_2
\ee
The symmetric and antisymmetric combination of the interaction terms define the two Landau parameters, which after normalization with respect to the density of states are
\be{landau}
F_0={{L}\over{2\pi v_s}}(f_0(k_F,k_F)+f_0(k_F,-k_F))={1\over g}-1\nn
F_1={{L}\over{2\pi v_s}}(f_0(k_F,k_F)-f_0(k_F,-k_F))=g-1
\ee
This gives the following relation \cite{Isakov96}
\be{rel}
1+F_1={1\over{1+F_0}}=g
\ee

It is of interest to relate this result to the condition of Galilean invariance, as expressed in the Fermi liquid formulation. This condition is written as\cite{Landau56},
\be{gal}
\int dk \,k \,n(k)=\int dk \,m\pdd{\ge(k)}{k} n(k) 
\ee
where $m$ is the (bare) mass of the fermions and the occupation numbers are treated as a continuous function of $k$. The equation states that the total momentum is conserved under the adiabatic turning on of the interaction.
Variation in the particle density gives
\be{nvar}
\int dk\, {k\over m}\gd n(k)=\int dk \pdd{\ge(k)}{k}\gd n(k)+{L\over{2\pi}}\iint dk dk' \,\pdd{f(k,k')}{k} n(k)\gd n(k')
\ee
where the last term is the result of treating $\ge(k)$ as a functional of $n(k)$.
Assuming this is valid for arbitrary variations $\gd n(k)$, we get the following relation
\be{km}
{k\over m}=\pdd{\ge(k)}{k}-{L\over 2\pi}\int dk'\,f(k,k')\pdd{n(k')}{k'}
\ee
For a filled Fermi sea the derivative of the particle density is
\be{dnk}
\pdd{n_0(k')}{k'}=\gd(k'+k_F)-\gd(k'-k_F)
\ee
and with $k=k_F$, Eq.\pref{km} gets the form
\be{kmf}
{k_F\over m}=\left.\pdd{\ge_0(k)}{k}\right |_{k_F}+{L\over 2\pi}(f_0(k_F,k_F)-f_0(k_F,-k_F))
\ee
We  make the following identifications
\be{iden}
v_F={k_F\over m}\,,\quad v_s=\left. \pdd{\ge_0(k)}{k}\right |_{k_F}\equiv {k_F\over {m^*}}
\ee
with $v_F$ as the Fermi velocity of the non-interacting fermions, $v_s$ as the Fermi quasiparticle velocity, and $m^*$ as the effective mass of the quasiparticles. This gives
\be{mm}
{m^*\over m}={v_F\over {v_s}}=1+F_1=g
\ee
If we further apply the identities
\be{identi}
v_s=\sqrt{v_J v_N}\,,\quad g=\sqrt{v_J/v_N}
\ee
we find
\be{vj}
v_J=v_F
\ee
which is consistent with the earlier result \pref{velpar}. Here it follows as consequence of Galilei invariance in Landau's Fermi liquid formulation, whereas the result in \pref{velpar} is a consequence of the corresponding symmetry of the two-particle interaction $V(k_1-k_2,q)$. The equality between $1+F_1$ and $(1+F_0)^{-1}$ in Eq.  \pref{rel}  can be seen as a consequence of the equality in the one dimensional system between the velocity of the quasiparticle excitations and the velocity of sound  $v_s$.
 
\section{Quantum Statistics}
A central element in the Fermi liquid theory is the assumption that the elementary excitations (quasiparticles) of the theory obey Fermi-Dirac statistics. This means that the entropy function has the same form as for the non-interacting (bare) particles,
\be{ent}
S= -\sum_k[n(k)\log n(k)+(1-n(k))\log(1-n(k))]
\ee
In the case discussed in the previous sections, this follows since the dressed particle field $\gf(x)$ is related to the original fermion field $\psi(x)$ by a unitary transformation. However, a further change of variable will now be introduced, which changes this relation. This is not done in the form of a transformation of the field operators, 
but rather by introducing new momentum variables, with a stronger repulsion between neighboring values than demanded by the Pauli exclusion. With $k$ as a discrete variable $k_i=2\pi n_i/L$, where the integer $n_i$ is assumed to increase monotonically with $i$,  a simple transformation which increases the exclusion effect in $k$-space is
\be{disc}
\gk_i=k_i+\gl {\pi\over L}\sum_j\sgn(k_i-k_j)\,,\quad i=1,2,...
\ee
where $\gl$ is a new, real parameter.
This leads to the following modified exclusion between the $\gk$ values 
\be{exc}
\gk_{i+1}=\gk_i +{2\pi\over L}(\gD n_i +\gl)
\ee
with $\gD n_i$ as a positive integer.

In the continuum limit ($L\to \infty$) the transformation \pref{disc} gets the form of a Bethe ansatz equation, which relates the pseudomomentum $\gk$ to the momentum $k$, and associated with this introduces a change of the particle density function, $n(k)\to\nu(\gk)$,
\be{momtrans}
&&\gk=k+\hf \gl\int dk'\,n(k')\,\sgn(k-k')\nn[2mm]
&&\nu(\gk)\,d\gk=n(k)\,dk
\ee
It follows directly that the two densities are related by
\be{dens}
\nu(\gk)=\frac{n(k)}{1+\gl n(k)}\,,\quad n(k)=\frac{\nu(\gk)}{1-\gl \nu(\gk)}
\ee
with $k$ and $\gk$ related as shown in \pref{momtrans}. The entropy expressed in the new variables is
\be{entro2}
S=-{L\over{2\pi}}\int_0^\infty d\gk[\nu\ln\nu-(1-\gl\nu)\ln (1-\gl\nu)\nn
+(1-(1+\gl)\nu))\ln (1-(1+\gl)\nu)]
\ee
which is recognizable as the entropy of system of particles with generalized  exclusion statistics \cite{Haldane91}.

However, one should note that at this point the transformation introduced above is only a change of variables. We cannot make any conclusion about the quantum statistics of the particles without considering what the transformation makes to the energy functional of the system.   
The point to show is that by choosing a particular value for the parameter $\gl$, the leading part of the quasiparticle interaction, defined in the previous section by the Landau parameters $F_0$ and $F_1$, is transformed to zero. This implies that the statistics defined by the new form of the entropy is not modified by a statistical interaction term. 

In order to show this we consider \pref{momtrans} as introducing a change of variables for the total energy, $E[n(k)]\to E[\nu(\gk]$. This redefines the quasiparticle energy and interaction 
\be{enin}
\tilde\ge(\gk)={{2\pi}\over L}\frac{\gd E}{\gd\nu(\gk)}\,,\quad \tilde f(\gk,\gk')={{4\pi^2}\over L^2}\frac{\gd^2E}{\gd\nu(\gk)\gd\nu(\gk')}
\ee
and the idea is to express these in terms of the previous functionals $\ge(k)$ and $f(k,k')$.
For the single particle energy the transformation gives
\be{entrans}
\te(\gk')={2\pi\over L}\int dk\, \frac{\gd n(k)}{\gd\nu(\gk')}\frac{\gd E}{\gd n(k)}=\int dk\, \frac{\gd n(k)}{\gd\nu(\gk')}\ge(k)
\ee
and for the interaction
\be{intfunc}
\tf(\gk'',\gk')&=&{{4\pi^2}\over L^2}\frac{\gd}{\gd\nu(\gk'')}\int dk\, \frac{\gd n(k)}{\gd\nu(\gk')}\frac{\gd E}{\gd n(k)}\nn
&=&{2\pi\over L}\int dk\,\frac{\gd^2 n(k)}{\gd\nu(\gk'')\gd\nu(\gk')}\ge(k)+\iint d\bar k\,dk 
\frac{\gd n(\bar k)}{\gd\nu(\gk'')}
\frac{\gd n(k)}{\gd\nu(\gk')} f(\brk,k)
\ee
After some tedious manipulations (see the Appendix), 
the following rather simple expressions are found for the transformation matrices,
\be{transmat}
\frac{\gd n(k)}{\gd\nu(\gk')}&=&\hf\dd{}{k}[(1+\gl n(k))\sgn(k-k')]\nn
\frac{\gd^2 n(k)}{\gd\nu(\gk'')\gd\nu(\gk')}&=&{1\over 4}\gl\ddtwo{}{k}[(1+\gl n(k))\sgn(k-k')\sgn(k-k'')]
\ee
where the pairs of variables $k', \gk'$ and $k'', \gk''$ are related by the transformation \pref{momtrans}.

The expression obtained for the energy is then the following
\be{entrans2}
\te(\gk')&=&\hf\int dk\, \ge(k)\dd{}{k}[(1+\gl n(k))\sgn(k-k')]\nn
&=& (1+\gl n(k'))\ge(k')+\hf \gl\int dk\, \ge(k)n'(k)\sgn(k-k')
\ee
with $n'(k)=dn/dk$. 

In the case of a filled Fermi sea, the particle density and its derivative are 
\be{dens2}
n_0(k)&=&\hf(\sgn(k+k_F)-\sgn(k-k_F))\nn
n'_0(k)&=&\gd(k+k_F)-\gd(k-k_F)
\ee
This gives for the pseudomomentum
\be{pseu}
\gk=k+\hf\gl\int_{-k_F}^{k_F} d\brk\,\sgn(k-\brk)=
\left\{
\begin{matrix} k+\gl k_F & \quad k>k_F \\ k(1+\gl) &\quad -k_F<k<k_F \\ k-\gl k_F &\quad k<-k_F
\end{matrix} 
\right.
\ee
and in particular $\gk_F=(1+\gl)k_F$. The transformed particle density then is
\be{trdens}
\nu_0(\gk)= {1\over{2(1+\gl)}}(\sgn(\gk+\gk_F)-\sgn(\gk-\gk_F))
\ee
Introducing this in the expression for the quasiparticle energy gives
\be{etil}
\te_0(\gk)&=&
\left\{
\begin{matrix}
\quad \ge_0(\gk-\frac{\gl}{1+\gl}\gk_F)&\quad \gk>\gk_F\\
\quad (1+\gl)\ge_0(\frac{\gk}{1+\gl})-\gl \ge_F&\quad |\gk|<\gk_F\\
\quad \ge_0(\gk+\frac{\gl}{1+\gl}\gk_F)&\quad \gk<-\gk_F
\end{matrix}
\right.
\ee
with  $\ge_F\equiv \ge_0(k_F)=\te_0(\gk_F)$.

For the interaction, we cite here only the results for variations around the filled Fermi sea, and refer to the Appendix for more details,
\be{int1}
&&\tf_{0}(\gk'',\gk')=-{\pi\over L}\gl(\ge'_0(k')(1+\gl n_0(k'))-\ge'_0(k'')(1+\gl n_0(k'')))\sgn(k'-k'')\nn
&&+{\pi\over 2L}\gl^2\ge'_0(k_F)\{\sgn(k''+k_F)\sgn(k'+k_F)+\sgn(k''-k_F)\sgn(k'-k_F)\}\nn
&& +(1+\gl n_0(k'))(1+\gl n_0(k''))f_0(k'',k')\nn
&&-\hf\gl(1+\gl n_0(k'))\{f_0(-k_F,k')\sgn(k''+k_F)-f_0(k_F,k')\sgn(k''-k_F)\}\nn
&& -\hf\gl(1+\gl n_0(k''))\{f_0(k'',-k_F)\sgn(k'+k_F)-f_0(k'',k_F)\sgn(k'-k_F)\}\nn
&&+{1\over 4}\gl^2\left [\left(\sgn(k''+k_F)\sgn(k'+k_F) +\sgn(k''-k_F)\sgn(k'-k_F)f_0(k_F,k_F)\right)\right . \nn
&&\quad\quad -\left(\left .\sgn(k''+k_F)\sgn(k'-k_F)+ \sgn(k''-k_F)+\sgn(k'-k_F)\right)f_0(k_F,-k_F)\right] \nn
\ee
In spite of the discontinuous behavior of the particle density $n(k)$ at the Fermi points, the function $\tf_{0}(\gk'',\gk')$ is continuous at these points, as one can check from the above expression. The values at the Fermi points are
\be{ftild}
\tf_0(\gk_F,\gk_F)&=&f_0(k_F,k_F)+(\gl+\hf\gl^2)(f_0(k_F,k_F)-f_0(k_F,-k_F))+{\pi\over L}\gl^2 v_s\nn
\tf_0(\gk_F,-\gk_F)&=&f_0(k_F,-k_F)-(\gl+\hf\gl^2)(f_0(k_F,k_F)-f_0(k_F,-k_F)+{2\pi\over L}v_s)
\ee
Assuming the new Landau parameters $\tilde F_0$ and $\tilde F_1$ to be normalized as in \pref{landau},   this implies the following relation to the original Landau parameters
\be{landpar}
\widetilde F_0&=&F_0-\gl\nn
\widetilde F_1&=&(1+\gl)^2(F_1+{\gl\over {1+\gl}})
\ee
Furthermore, the relation \pref{rel} between $F_0$ and $F_1$ gives the following relation between $\tilde F_0$ and $\tilde F_1$,
 \be{relnew}
 (1+\frac{\tilde F_0}{1+\gl})=( 1+\frac{\tilde F_1}{1+\gl})^{-1}
 \ee
 
Assuming now that the value of the parameter $\gl$ is specified as
\be{defgl}
\gl=\gl_1+\gl_2={1\over g}-1
\ee
and using the values earlier found for $F_0$ and $F_1$ in \pref{rel},  we find that both the new Landau parameters vanish,
\be{van}
\widetilde F_0=\widetilde F_1=0
\ee
This means that both $\tf_0(\gk_F,\gk_F)$ and $\tf_0(\gk_F,-\gk_F)$ vanish, and therefore that the interaction in the low energy regime is weak, in the sense
\be{lim3}
\lim_{|\gk''|\to k_F}\,\lim_{|\gk'|\to k_F} \tf_0(\gk'',\gk')=0
\ee

The conclusion is thus that the interactions of the one-dimensional fermion system effectively change the particle statistics, and make the system appear as a weakly interacting system of (quasi)particles with generalized statistics. The modified exclusion parameter is given by $1+\gl=1/g$,
which agrees with the value of the statistics parameter in Refs.~\onlinecite{WuYu95, Isakov98,WuYu01}.

\section{The low energy limit}
As a consistency check we briefly return to the linearized low energy approximation, with the Hamiltonian described by \pref{hamdiag}. In this approximation the interaction terms $\tilde f(\gk'',\gk')$ are negligible, and the energy function $\te(\gk)$ is essentially independent of the particle density and can be approximated by $\te_0(\gk)$. The point to check is whether the energies $E=\int d\gk\, \nu(\gk)\,\te_0(\gk)$, with the particle distributions specified by \pref{disc}, coincide with the eigenvalues of the Hamiltonian \pref{hamdiag}.

We first assume the particle distribution to be without holes, where all momentum states $k_i$ are filled between a minimum value $k_{min}$, close to the Fermi point $-k_F$, and a maximum value $k_{max}$, close to $k_F$. $k_{min}$ and $k_{max}$ are then related to the particle numbers $N$ and $J$ in the following way 
\be{nj}
k_{max}=k_F+{\pi\over L}(N+J)\,,\quad k_{min}=-k_F-{\pi\over L}(N-J)
\ee
and the relation \pref{momtrans} between $\gk$ and $k$ simplifies to
\be{simp}
\gk=(1+\gl) k-{\pi\over L}\gl J
\ee
The integrated energy of the system is
 \be{cont}
E&=&{L\over{2\pi}}\int_{\gk_{min}}^{\gk_{max}}\te_0(\gk) \nu_0(\gk)d\gk\nn
&=&{L\over{2\pi}}{1\over{1+\gl}}\left(\int_{-\gk_F}^{\gk_F}\te_0(\gk)d\gk+\int_{\gk_{min}}^{-\gk_F}\te_0(\gk)d\gk
+\int_{\gk_F}^{\gk_{max}}\te_0(\gk)d\gk\right)\nn
&=&E_0+{L\over{2\pi}}{1\over{1+\gl}}\left(\int_{\gk_{min}}^{-\gk_F}(\ge_F-v_s(\gk+\gk_F))d\gk
+\int_{\gk_F}^{\gk_{max}}(\ge_F+v_s(\gk-\gk_F))d\gk\right)\nn
&=&E_0+\ge_F N+ {\pi\over 2L} v_s((1+\gl)N^2+{1\over {1+\gl}} J^2)
\ee
where $E_0$ is the energy of the filled Fermi sea, and where $\te_0(\gk)$ has been expanded to linear order in $\gk$ in the intervals close to the Fermi points. The result agrees with the $N$ and $J$ dependent terms of the Hamiltonian \pref{hamdiag}. The two first terms in \pref{cont} can be absorbed in a redefinition of the zero point of the energy and by introducing a chemical potential, which effectively makes the Fermi sea the ground state of the system. The bosonic excitation term in \pref{hamdiag} corresponds to the particle-hole excitations of the fermionic system, which so far has not been included.

In Fig.~1 the situation is illustrated for a small system with quadratic dispersion, $\te_0=\hf a \gk^2$, $a=v_s/\gk_F$, and with statistics parameter $\gl=0.5$. The single particle energies $\te_0$ are shown as functions of $k$ in two cases, the first one with a filled Fermi sea, with $N_0=23$ particles and $N=J=0$. In the second case two particles have been added in the lowest available states close to the Fermi point $k_F$, so that $N=J=2$. The figure demonstrates how the effects of the modified Pauli exclusion are transmitted through the Fermi sea. This produces effectively a shift of the energy curve along the $k$ axis. Thus, the insertion of new particles near the Fermi point $k_F$ will affect  the values of $\gk$, and thus of $\te_0$, not only near the Fermi point $k_F$, but also near $-k_F$. It is straight forward to to show that this will effectively move a part of any new fermion charge inserted at $k_F$, so that the fraction $(1+g)/2$ of the charge remains at $k_F$, while the fraction $(1-g)/2$ is moved to $-k_F$.
\begin{figure}[h]
\begin{center}
\includegraphics[width=11.0cm]{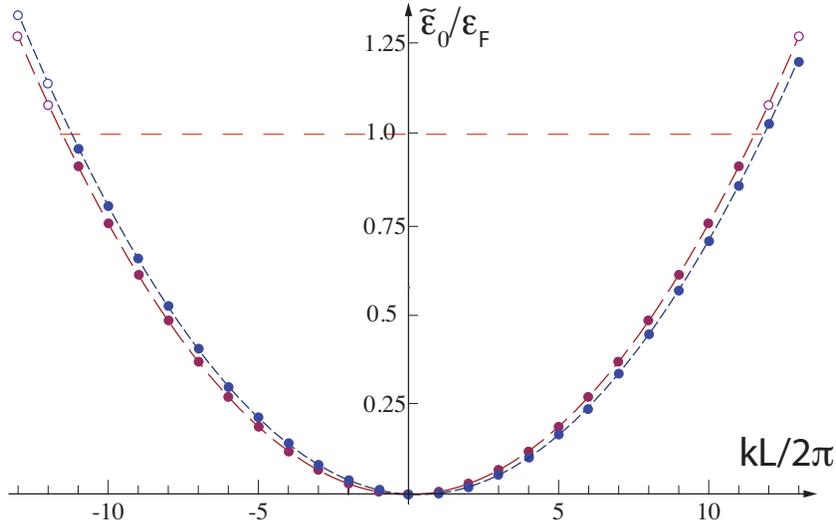}
\end{center}
\caption{\small The single particle energy $\te_0$, shown as function of the momentum variable $k$, with quadratic dispersion and with statistics parameter $\gl=0.5$. Filled circles correspond to occupied states and open circles to unoccupied states. Two cases are shown. The first case (red circles, long dashed curve) corresponds to a filled Fermi sea, representing the ground state with $N_0=23$ particles. In the second case (blue circles, short dashed curve) two particles are added in the lowest available momentum states close to the Fermi point $k_F$.  This corresponds to the case $N=J=2$ .The generalized Pauli exclusion between the particles gives rise to a relative shift of the two curves along the $k$ axis.  \label{energy}}
\end{figure}

Particle-hole excitations can be introduced by changing the (discrete) momenta $k_i$ of the occupied states in the following way. 
\be{disk}
 k_i=k_i^0 +\gD k_i=(2\pi/L) (i+n_i)\,, \quad\quad i=i_{min},i_{min}+1,...,i_{\max}
 \ee
where $k_i^0$ gives the momentum values of the occupied states without holes, and  $n_i$ are integers, which introduce holes in the distribution. We assume $n_{i+1} \geq n_i $ to make the ordering of $k_i$ with respect to $i$ to be unchanged when introducing the holes. For the pseudomomenta $\gk_i$ we  get a similar change in the values. This is a direct consequence of the transformation formula \pref{disc}, 
\be{disk2}
\gk_i&=&k_i+\gl{\pi\over L}\sum_{j=i_{min}}^{i_{max}} \sgn(k_i-k_j)\nn
&=& 2\pi/L)( i+n_i+\hf\gl\sum_{j=i_{min}}^{i_{max}}  \sgn(i-j))\nn
&\equiv&\gk_i^0+\gD k_i
\ee
with $\gD k_i=(2\pi/L) n_i$. The shifts are thus the same as for the momenta $k_i$, which means that they are independent of the statistics parameter $\gl$.

With the excitations restricted to the neighborhoods of the Fermi points, linearization of the energy as function of momentum can be made, which gives
\be{nysum}
\sum_i \te_0(\gk_i)=\sum_i \te_0(\gk_i^0)+v_s\sum_i|\gD k_i|
\ee
The excitation term is the same as for free fermions, although with $v_s$ as the effective Fermi velocity. It corresponds precisely to the bosonic excitation term in \pref{hamdiag}, when the boson frequency is linearized in $q$, $\gw_q\approx v_s |q|$. Thus the expression for the full energy, given as a sum over single particle energies $\te_0(\gk_i)$ reproduces precisely the energy eigenvalues of the Hamiltonian \pref{hamdiag} within this approximation.

\section{Concluding remarks}
The standard approach to study the interacting one-dimensional Fermi system is based on the use of the bosonization technique. However, to study properties of the elementary charged excitations, other methods may be more convenient.  We have here applied the  Fermi liquid theory of Landau to study the quantum statistics of the "dressed fermions" of the interacting theory. These particles are identified as the Landau type of quasiparticles, and they are well defined due to the adiabatic mapping which exists between the (low energy sector of the) non-interacting and the interacting theory.

For excitations close to the Fermi points, the unitary transformation between the non-interacting and the interacting system is explicitly known and has a simple form when expressed in the bosonized variables. We have used this to rewrite the low energy Hamiltonian in terms of dressed fermion fields, and to show that the corresponding excitations are fractionally charged  A more detailed study of this effect has been performed in a previous publication \cite{LeinaasHorsdal09}. The Hamiltonian, in this form, has furthermore been used to determine the quasiparticle energy and two-body interaction, defined as functional derivatives of the full energy of the system. 
By a change of momentum variables in the functional derivatives, the interaction terms have been shown to vanish for momenta close to the Fermi points. This formally gives the description the form of a free theory. However, as shown by the transformed form of the entropy function, the quasiparticles obey a generalized type of Pauli exclusion.

The two-body interaction, after the momentum transformation, vanishes at the Fermi points. Away from these points, however there are non-vanishing corrections, as shown by the derived expressions. Since the main part of the interaction has been absorbed in the change of statistics, these contributions are small and can in principle be treated perturbatively. 

\vskip5mm 
\noi
{\bf Acknowledgment}\\
I would like to thank Hans Hansson for valuable comments to this work. The work has been financially supported by the Norwegian Research Council.
\\

\appendix
\noi
{\bf Appendix: Doing the functional derivatives}\\
In order to determine $\te$ and $\tf$ we need to evaluate the functional derivatives
$\frac{\gd n(k)}{\gd\nu(\gk')}$ and $\quad \frac{\gd^2 n(k)}{\gd\nu(\gk'')\gd\nu(\gk')}$ 
as shown in Eqs.~\pref{entrans} and \pref{intfunc}. $n(k)$ can be viewed as a functional of $\nu$ (and a function of $k$), as specified in Eq.~\pref{momtrans},
\be{func}
n(k)=\frac{\nu(\gk)}{1-\gl\nu(\gk)}\equiv F[\nu(\gk)]
\ee
One should however note that with $k$ fixed, the corresponding variable $\gk$ will be a functional of $\nu$, as shown by the relation
\be{kappa}
\gk=k+\hf\gl\int d\gk' \nu(\gk')\sgn(\gk-\gk')
\ee
We write the variation of the functional $F$ in the following way
\be{fvar}
\gd F[\nu(\gk)]=F[\nu(\gk+\gd\gk)+\gd \nu(\gk+\gd\gk)]-F[\nu(\gk)]
\ee
with
\be{varnu}
\gd \nu(\gk)=\ge_1\gd(\gk-\gk')+\ge_2 \gd(\gk-\gk'')
\ee
When evaluating $\gd F$ to second order in $\ge_1$ and $\ge_2$, the term proportional to $\ge_1$ will give the functional derivative with respect to $\gk'$ and the second order term proportional to $\ge_1\ge_2$ will give the double derivative with respect to $\gk'$ and $\gk''$.

We begin by evaluating (to second order) the variation in $\gk$ induced by the variation in $\nu$,
\be{varkap}
\gd\gk=\hf\gl\int d\bar k \left[(\nu(\bar k)+\gd\nu(\bar k))\sgn(\gk+\gd\gk-\bar k)-\nu(\bar k)\sgn(\gk-\bar k)\right]
\ee
We have
\be{sgn}
\sgn(\gk+\gd\gk-\bar k)=\sgn(\gk-\bar k)+2(\gq(\gk+\gd\gk-\bar \gk)-\gq(\gk-\bar \gk))
\ee
with $\gq(\gk-\bar \gk)$ as the Heaviside step function, and using the expansion
\be{expan}\nu(\bar\gk)=\nu(\gk)+\nu'(\gk)(\bar\gk-\gk)...
\ee
we find the following implicit expression for $\gd\gk$,
\be{dk}
\gd\gk=\hf\gl\left[\int d\bar\gk\,\gd\nu(\bar\gk)\sgn(\gk-\bar\gk)+2(\nu(\gk)+\gd\nu(\gk))\gd\gk+\nu'(\gk)\gd\gk^2\right]+ {\rm higher\;\; order}\nn
\ee
From this the first order contribution is found to be
\be{kone}
\gd\gk_1&=&\frac{\gl/2}{1-\gl\nu(\gk)}\int d\bar\gk \gd\nu(\bar\gk)\sgn(\gk-\bar\gk)\nn
&=&\frac{\gl/2}{1-\gl\nu(\gk)}(\ge_1\sgn(\gk-\gk')+\ge_2\sgn(\gk-\gk''))
\ee
and including explicitly only the term proportional to $\ge_1\ge_2$ we find for the second order term
\be{ktwo}
\gd\gk_2&=&2\ge_1\ge_2\left[(\frac{\gl/2}{1-\gl\nu(\gk)})^2(\gd(\gk-\gk')\sgn(\gk-\gk'')+\gd(\gk-\gk'')\sgn(\gk-\gk')\right .\nn
&& \quad\quad\quad\left .+(\frac{\gl/2}{1-\gl\nu(\gk)})^3\nu'(\gk)\sgn(\gk-\gk')\sgn(\gk-\gk'')\right]+...
\ee

We consider now the expansion of $\gd F[\nu(\gk)]$ in powers of $\ge_1$ and $\ge_2$, starting with the expression
\be{fexp}
\gd F[\nu(\gk)]= {1\over \gl}\left(\frac{1}{1-\gl(\nu(\gk+\gd\gk)+\gd\nu(\gk+\gd\gk)}-\frac{1}{1-\gl\nu(\gk)}\right)
\ee
We leave out the intermediate steps in deriving the expansion and give the expression for the coefficient proportional to $\ge_1$, which determines the functional derivative of $n(k)$ with respect to $\nu(\gk')$,
\be{first}
\frac{\gd n(k)}{\gd\nu(\gk')}=\frac{1}{(1-\gl\nu(\gk))^2}\left(\gd(\gk-\gk')+\frac{\gl/2}{(1-\gl\nu(\gk)}\nu'(\gk)\sgn(\gk-\gk')\right)
\ee
From the term proportional to $\ge_1\ge_2$ we derive the following expression
\be{second}
\frac{\gd^2 n(k)}{\gd\nu(\gk'')\gd\nu(\gk')}&=& {1\over 4}\left(\frac{\gl^2}{(1-\gl\nu(\gk)^4}\nu''(\gk)+3\frac{\gl^3}{(1-\gl\nu(\gk)^5}\nu(\gk)'^2\right)\sgn(\gk-\gk')\sgn(\gk-\gk'')\nn
&+&{3\over 2}\frac{\gl^2}{(1-\gl\nu(\gk)^4}\nu'(\gk)(\sgn(\gk-\gk')\gd(\gk-\gk'')+\gd(\gk-\gk')\sgn(\gk-\gk''))\nn
&+&\hf\frac{\gl}{(1-\gl\nu(\gk)^3}(\sgn(\gk-\gk')\gd'(\gk-\gk'')+\gd'(\gk-\gk')\sgn(\gk-\gk''))\nn
&&+2\frac{\gl}{(1-\gl\nu(\gk)^3}\gd(\gk-\gk')\gd(\gk-\gk'')
\ee

Finally we rewrite these in terms of the variable $n(k)$, by use of the identity
\be{ident}
\frac{1}{1-\gl\nu(\gk)}=1+\gl n(k)
\ee
Leaving out also here the intermediate steps we find the following expressions
\be{finalone}
\frac{\gd n(k)}{\gd\nu(\gk')}&=&(1+\gl n(k))^2(\frac{1}{1+\gl n(k)}\gd(k-k')+\frac{\gl/2}{(1+\gl n(k))^2}n'(k)\sgn(k-k'))\nn
&=&\hf\dd{}{k}[(1+\gl n(k)\sgn(k-k')]
\ee
and
\be{finaltwo}
\frac{\gd^2 n(k)}{\gd\nu(\gk'')\gd\nu(\gk')}&=& {1\over 4}\gl^2 n''(k)\sgn(k-k')\sgn(k-k'')\nn
&&+\gl^2n'(k)(\sgn(k-k')\gd(k-k'')+\gd(k-k')\sgn(k-k'')\nn
&&+\hf\gl(1+\gl n(k))(\sgn(k-k')\gd'(k-k'')+\gd'(k-k')\sgn(k-k'')\nn
&&+2\gl(1+\gl n(k))\gd(k-k')\gd'(k-k'')\nn
&=&{1\over 4}\gl\ddtwo{}{k}[(1+\gl n(k))\sgn(k-k')\sgn(k-k'')]
\ee

The continuum form of the expression for the variation in the total energy \pref{enpart2} we write as
\be{enpart2}
\gd E={L\over{2\pi}}\int dk \ge(k) \gd n(k)+{L^2\over{8\pi^2}}\iint dk dk' f(k,k')\gd n(k)\gd n(k')
\ee
which gives
\be{enin2}
\ge(k)={{2\pi}\over L}\frac{\gd E}{\gd n(k)}\,,\quad f(k,k')={{4\pi^2}\over L^2}\frac{\gd^2E}{\gd n(k)\gd n(k')}
\ee
The energy function after the change of variables then is
\be{efunc}
\te(\gk')&=&{{2\pi}\over L}\frac{\gd E}{\gd \nu(\gk')}\nn
&=& \int dk\, \frac{\gd n(k)}{\gd\nu(\gk')}\ge(k)\nn
&=&\hf\int dk \ge(k)\dd{}{k}[(1+\gl n(k))\sgn(k-k')]\nn
&=&(1+\gl n(k'))\ge(k')+\hf\gl\int dk \ge(k) n'(k)\sgn(k-k')
\ee
The interaction function it is convenient to separate in two parts
\be{int2}
\tf(\gk'',\gk)&=& {{4\pi^2}\over L^2}\frac{\gd^2E}{\gd \nu(\gk)\gd \nu(\gk')}\nn
&=&{2\pi\over L}\int dk \frac{\gd^2 n(k)}{\gd \nu(\gk'')\gd \nu(\gk')} \ge(k)+\iint d\bar k dk \frac{\gd n(k)}{\gd \nu(\gk')} \frac{\gd n(k)}{\gd \nu(\gk'')}f(\bar k, k)\nn[2mm]
&=& \tf_A(\gk'',\gk')+\tf_B(\gk'',\gk')
\ee
For the first term we find
\be{inta}
\tf_A(\gk'',\gk')&=&{\pi\over {2L}}\gl\int dk \ge(k)\ddtwo{}{k}[(1+\gl n(k))\sgn(k-k')\sgn(k-k'')]\nn
&=&-{\pi\over {L}}\gl[\ge'(k')(1+\gl n(k'))-\ge'(k'')(1+\gl n(k''))]\sgn(k'-k'')\nn
&&-{\pi\over {2L}}\gl^2\int dk\, \ge'(k)\, n'(k)\,\sgn(k-k')\sgn(k-k'')
\ee
and for the second term
\be{intb}
\tf_B(\gk'',\gk')&=&{1\over 4}\iint d\bar k\, dk\, \dd{}{\bar k}[(1+\gl n(\bar k))\sgn(\bar k-k'')]
\dd{}{k}[(1+\gl n(k))\sgn(k-k')] f(\bar k,k)\nn
&=&(1+\gl n(k''))(1+\gl n(k'))f(k'',k')\nn
&& + \hf\gl(1+\gl n(k''))\int dk\,  n'(k)\,\sgn(k-k')f(k'',k)\nn
&& + \hf\gl(1+\gl n(k'))\int dk\,  n'(k)\,\sgn(k-k'')f(k,k')\nn
&&+{1\over 4}\gl^2\iint d\bar k dk n'(\bar k) n'(k)\sgn(\bar k-k'')\sgn(k-k')f(\bar k,k)
\ee
In the expressions given above the pair $k'$ and $\gk'$, as well as $k''$ and $\gk''$, are related by Eq.~\pref{kappa}. 

The expressions for the energy $\te_0(\gk)$ and for the interaction $\tf_0(\gk'',\gk')$ are finally found by introducing the density $n_0(k)$ and its derivative $n'_0(k)$, as given by Eq.~ \pref{dens2}, in the above expressions \pref{efunc}, \pref{inta}, and \pref{intb}. The results are displayed in Eqs.~\pref{etil}  and \pref{int1} in the main text. 


\end{document}